\documentclass[12pt]{iopart}
\usepackage{graphicx}

\begin{document}
\title{Spin-flip scattering in chiral induced spin selectivity using the Riccati equation}

\author{D N\"urenberg, H Zacharias}
\address{Center for Soft Nanoscience, Westf\"alische-Wilhelms Universit\"at, 48149 M\"unster, Germany}

\begin{abstract} 

The chiral induced spin selectivity (CISS) in layers of helical molecules gained considerable attention in the emerging field of spintronics, because the effect enables spin-filter devices under ambient conditions. Several theoretical studies have been carried out to explain this effect on a microscopic scale, but the origin of the effect is still controversial. In particular the role of spin-flip scattering during electron transport is an open issue. In this study we describe the electron and spin transport by rate equations including spin-dependent losses and spin-flip scattering. We reduce the problem to the solution of the Riccati differential equation to obtain analytical solutions. The results allow to determine and interpret the strength and scalability of CISS based spin-filters  from experimental data or quantum mechanical models. For the helical systems studied experimentally so far it turns out that spin-flip scattering plays a minor role.
 
\end{abstract}


\maketitle

\section{Introduction}
In recent years it has been shown that electron transmission through monolayers of helical molecules adsorbed on various substrates show a preferred longitudinal spin orientation. This spin orientation depends on the enatiomeric chiral orientation of the molecules \cite{Kiran:2016,Kettner:2018}. This so-called chirally induced spin selectivity (CISS) effect \cite{Naaman:2015} attractes presently considerable attention as a route towards room temperature spintronic devices \cite{Mathew:2014, Koplovitz:2017}. Spin polarised currents caused by chiral molecules may further advance the efficiency of water splitting \cite{Mtangi:2015,Mtangi:2017} and are debated  to play a role in biological systems \cite{Michaeli:2016}. First experiments by Naaman and coworkers \cite{Ray:1999, Ray:2006} measured intensity asymmetries in the photoemission from gold substrates coated with films of chiral molecules when applying left and right circularly polarised photons. Gold is known to produce spin-polarised photoelectrons upon excitation with circularly polarised UV radiation and the asymmetry may consequently be caused by spin-dependent losses in the film. Other experiments used electron conduction measurements through chiral molecules contacted to ferromagnetic metals and find asymmetries in the conduction with respect to the magnetisation direction of the substrate \cite{Xie:2011}. These measurement techniques give only an indirect hint at spin polarised currents and are hardly sensitive to spin-flips. This lack was overcome by direct measurements of the electron spin orientation by Mott schattering after transmission through a chiral film of double stranded DNA \cite{Goehler:2011}.\\
Considerable theoretical effort has been taken to explain the spin selective electron transport in chiral molecules - yet several models are competing to explain the effect. These models calculate either spin-dependent electron scattering at a microscopic level of a model molecule \cite{Yeganeh:2009}, or conduction through a such model \cite{Gutierrez:2012,Guo:2012b}. Conduction of electrons along the backbone of a molecule can be calculated by choosing a Hamiltonian which modeles the molecule and then use a spin-dependent Landauer formalism to calculate the transport in steps along the molecule. To model the Hamiltonian, the molecule was approximated by binding sites along a helix \cite{Gutierrez:2012,Gutierrez:2013} and a double helix \cite{Guo:2012b}, respectively. Both models result in an increasing spin polarisation of the current with an increasing length of the molecule.  Sun and Guo \cite{Guo:2012b} suggest the necessity of a dephasing of the wave function by inelastic scattering at each step for a significant spin polarising effect, but also find a saturation of the spin polarisation with the molecular length depending on the dephasing.\\
The average spin orientation of electrons  transmitted through the molecular film are affected by spin-dependent back-scattering or absorption, i.e. losses, and forward scattering with spin-flips. The role of the dephasing and of spin-flips are still controversial. The spin-flipping has not been addressed in the interpretation of experiments, because a description applicable to the data is missing. For the interpretation of the experiments an analytical model which can distinguish between spin-dependent losses and spin-flips would be useful. 
There are analytical desciptions for spin-dependent conduction through helical molecules based on the binding site model \cite{Matityahu:2016} or a 2D harmonic potential around a helical wire \cite{Michaeli:2015, Michaeli:2017}. However, for the identification of spin-flipping in the signal a more general approach would be helpful. Eremko and Loktev \cite{Eremko:2013} provide an analytical solution for a pure spin-state traveling through a helical potential and find an oscillatory behavior of the spin polarisation with the molecular length. As the authors pointed out, the pure spin state is not realized in current experiments. A suitable method to treat a mixed state is the density matrix formalism, which can describe scattering by scattering matrices. Blum and Thompson \cite{Blum:1989} used this formalism to investigate the spin-dependent forward scattering of an unpolarised ensemble - mixed state electrons - at molecular targets in the gas phase. They predicted from symmetry arguments that by planar molecules, no spin-dependent losses are possible in forward scattering, but spin-flips in longitudinal direction can occur if the direction of propagation is perpendicular to the plane of the molecule. In contrast, no spin-flips can occur in chiral molecules with $C_2$ symmetry, though a spin-selective loss can be present. These conciderations were made for the gas phase, on the other hand an assembly on a surface is more complicated since it breaks the $C_2$ symmetry of a helix for example. Such a break may play a role in the CISS effect, as Michaeli and Naaman predicted that an additional dipole moment along the propagation direction is even mandatory for large CISS effects \cite{Michaeli:2015, Michaeli:2017}. Approaches to calculate scattering at helical potentials by Mujica and coworkers result in very low longitudinal spin selectivity in single \cite{Yeganeh:2009} forward scattering, but for incoherent multiple forward scattering a spin polarisation up to 30\% was predicted \cite{Medina:2012, Rosenberg:2013}.\\
There exists a general approach  \cite{Farago:1971} to describe the evolution of a quantity of electron spins analog to the Stokes formalism in statistical polarisation optics, where the state of polarisation is fully described by a four component vector and each propagation step can be described by a 4x4 Matrix. CISS is sensitive to the longitudinal component of the electron spin, so the parameters of interest are spin-dependent losses and spin-flips between the spin-up and spin-down channel.  A very intuitive way to describe the change of spin polarisation in the longitudinal direction by forward-scattering was found by Fandreyer et al. \cite{Fandreyer:1990}.  A pair of differential equations describe the change of the expectation values for the number of spin-up, $N_1$, and spin-down electrons, $N_2$, as coupled rate equations. This approach is convenient for our purposes, because the coupling parameters directly describe the losses and spin-flips for each spin channel in longitudinal direction, i.e. the parameters are equal to the inverse of the mean free path of the corresponding process. The electron spin polarisation then can be obtained by its definition
\begin{eqnarray}
	P=\frac{N_1-N_2}{N_1+N_2}.
	\label{eq:DefSpinPol}
\end{eqnarray}
Although spin-flips are included in the rate equations, in their following discussion Fandreyer et al. could omit them for describing the propagation of a polarised electron beam through a vapor of chiral molecules as the detector cone in the related experiment \cite{Campbell:1985} was sufficiently small so that any scattering into the detector seemed unlikely.  In contrast to this gas phase experiment, for a transmission measurement through a film more scattering angles into the forward direction will contribute to the signal. Accordingly, we keep the spin-flip components and find that the evolution of the spin polarisation is mathematically equivalent to the Riccati equation. This equation has been used previously to describe related problems like the propagation of polarised light through optical media \cite{Azzam:1972} or the evolution of coherence in qubits \cite{Gardas:2010}. In this paper we will show analytical solutions and use them to obtain parameters for spin-flipping and spin-dependent loss probabilities from experimental data and from numerical results.
\section{Spin-flip and attenuation processes}
\subsection{Spin-dependent transport}
\label{chap:Riccati}
We define $N_1(x)$ as the expectation value of the number of electrons with spin up, i.e. spin parallel to the propagation direction along $x$, and $N_2(x)$ with the respective spin down or antiparallel spin. We then assume spin-dependent loss probabilities $\alpha_1(x)$ and $\alpha_2(x)$ per length unit while the electrons are traveling through the chiral layer. These losses may be caused by absorption in the molecule or by scattering out of the detector angle. We further assume a probability $f_1(x)$ per length for the electron to flip its spin from up to down and $f_2(x)$ vice versa. The physical origin of these contributions may be (multiple) scattering events with random spin-flip or spin-flip towards a preferred spin state. The inverse of the parameters $\alpha_1$, $\alpha_2$, $f_1$, and $f_2$ denotes the mean free path of the respective process. For this configuration we can set up coupled differential equations \cite{Fandreyer:1990}
\numparts
\begin{eqnarray}
	\frac{\textrm{d}N_1(x)}{\textrm{d}x}= -(\alpha_1(x)+f_1(x)) N_1(x) + f_2(x) N_2(x)  \label{eq:diff1}\\
	\frac{\textrm{d}N_2(x)}{\textrm{d}x}= -(\alpha_2(x)+f_2(x)) N_2(x) + f_1(x) N_1(x). \label{eq:diff2}
\end{eqnarray}
\endnumparts
In the following we abbreviate spatial derivatives in the form $\frac{\textrm{d}N_i(x)}{\textrm{d}x}=N_i'(x)$, and for convenience we omit to write the $x$ dependency explicitly, e.g., $N_i(x)=N_i$.  In an extension of the model of Fandreyer et al. \cite{Fandreyer:1990} we retain the spin-flip contributions $f_1$ and $f_2$. The differential equations (\ref{eq:diff1}), (\ref{eq:diff2}) can be rearranged to obtain a differential equation for the spin polarisation $P$ which is independent from the number of electrons in both spin states $N=N_1+N_2$. Using $\Delta N=N_1-N_2 $ for the difference of the numbers of electrons in both states and with
\numparts
\begin{eqnarray}
	N_1=(N+\Delta N)/2\\
	N_2=(N-\Delta N)/2,
\end{eqnarray}
\endnumparts
we arrive at
\numparts
\begin{eqnarray}
	N'+\Delta N'= -(\alpha_1+f_1)(N+\Delta N)+f_2 (N-\Delta N)\\
	N'-\Delta N'= -(\alpha_2+f_2)(N-\Delta N)+f_1 (N+\Delta N).
\end{eqnarray}
\endnumparts
We rearrange these equations to
\numparts
\begin{eqnarray}
	N'+\Delta N'= (-\alpha_1-f_1+f_2)N-(\alpha_1+f_1+f_2)\Delta N \label{eq:diff1a}\\
	N'-\Delta N'= (-\alpha_2-f_2+f_1)N+(\alpha_2+f_2+f_1)\Delta N. \label{eq:diff2a}
\end{eqnarray}
\endnumparts
Sum and difference of both equations yield
\numparts
\begin{eqnarray}
	2N'=-(\alpha_1+\alpha_2)N+(\alpha_2-\alpha_1) \Delta N\\
	2\Delta N'=(-\alpha_1+\alpha_2-2f_1+2f_2)N+(-\alpha_1-\alpha_2-2f_1-2f_2) \Delta N. 
\end{eqnarray}
\endnumparts
We abbreviate $\alpha=\alpha_1+\alpha_2$, which corresponds to twice the loss coefficent for unpolarised electrons, and   $f=f_1+f_2$, which is accordingly twice the spin-flip scattering probability in both spin directions for unpolarised electrons. $\Delta \alpha=\alpha_1-\alpha_2$ denotes the difference between the probabilities for electron loss in each spin state and $\Delta f= f_1 -f_2$ is the difference in the probabilities for spin-flip forward scattering. Expressed with these coefficients the differential equations are
\numparts
\begin{eqnarray}
	2N'=-\alpha N-\Delta \alpha \Delta N\\
	2\Delta N'=-(\Delta \alpha+2\Delta f)N-(\alpha+2f) \Delta N. 
\end{eqnarray}
\endnumparts
We can replace $\Delta N$ by $PN$, because $\Delta N/N=(N_1-N_2)/(N_1+N_2)=P$ defines the spin polarisation given by equation (\ref{eq:DefSpinPol}). The change of the number of electrons in the ensemble with the propagation length $x$ is then given by
\numparts
\begin{eqnarray}
	N'=-(\alpha+\Delta \alpha P) N/2. 
	  \label{eq:N}
\end{eqnarray}
The change of the difference of the numbers of electrons in both spin channels can be rearranged to
\begin{eqnarray}
	2(PN)'= 2(PN'+NP')=-(\Delta \alpha+2\Delta f)N-(\alpha+2f) PN, 		\label{eq:diffP2}. 
\end{eqnarray}
\endnumparts
and by combination of equations (\ref{eq:N}) and (\ref{eq:diffP2}) one obtains
\begin{eqnarray}
	P[-(\alpha+\Delta \alpha P) N]+2NP'= -(\Delta \alpha+2\Delta f)N-(\alpha+2f) PN.
\end{eqnarray}
For physical reasons we are only interested in situations where $N$ is non-zero. Thus the change of the spin polarisation with propagation is given by
\begin{eqnarray}
	 P'= \frac{\Delta \alpha}{2} P^2 -fP-\frac{\Delta \alpha}{2}-\Delta f.
	\label{eq:diffPend}
\end{eqnarray}
This nonlinear equation is independent of the number of electrons $N$. It describes the change of the spin polarisation as a function of propagation distance through the organic film due to loss and spin-flip processes. The equation is equivalent to Riccatis equation which is also used, e.g., to model the propagation of light in thin films for ellipsometry \cite{Azzam:1972} or to describe the loss of coherence in qubits with time \cite{Gardas:2010}.\\
With the substitution
\begin{eqnarray}
	P=P^*+1/\widetilde{P}
	\label{eq:Pubs}
\end{eqnarray} 
the Riccati equation can be transformed into the linear differential equation \cite{BronsteinDeu}
\begin{eqnarray}
	\widetilde{P}'=(f-\Delta \alpha P^*)\widetilde{P}-\Delta \alpha/2,
	\label{eq:redRiccati}
\end{eqnarray}
where $P^*$ is a special solution to the Riccati equation and $\tilde{P}$ is solution to equation (\ref{eq:redRiccati}).  With this substitution it is much easier to find a function for the spin polarisation, as soon as a special solution is known. Such a special solution can, e.g., be a fixed spin polarisation which stays constant during propagation because polarising and depolarising effects cancel out at this value. In this case the whole dynamic part is governed by the reduced equation  (\ref{eq:redRiccati}). We search for such special solutions, namely equilibrium points, in the next paragraph.
\subsection{Equilibrium points}
\label{chap:Trivial}
Up to here, the derivation is not restricted to coefficients which are constant along $x$. In the following we obtain analytical solutions for constant values of $\Delta \alpha$, $f$, and $\Delta f$. In the first place we look for spin polarisations $P^*$ which stay constant with propagation along $x$. These points with $P'=0$ are called equilibrium points and are useful special solutions $P^*$ to solve the Riccati equation with the substition approach given by equations (\ref{eq:Pubs}) and (\ref{eq:redRiccati}). There are stable and  unstable equilibrium points $P^*$ in the sense that any spin polarisation around a stable point $P^*$ tend towards $P^*$ with propagation along $x$ and instable ones tend away from $P^*$. This can be checked by the criterion for stability $\mathrm{d}P'/\mathrm{d}P|_{P^*}<0$. In this case the spin polarisation reaches the maximum at $\pm1$  or a state with equivalent polarising and depolarising effects.\\
To avoid divisions by zero one has to treat some cases separately. 
If $\Delta \alpha$, $\Delta f$ and $f$ are zero, no changes are made to the spin polarisation and every constant spin polarisation fulfills the differential equation (\ref{eq:diffPend}). Without spin-dependent attenuation, i.e. $\Delta \alpha =0$, but with spin-flip processes $ f\neq 0$ it follows
\begin{eqnarray}
	P^*=-\Delta f/f
	\label{eq:staticf}
\end{eqnarray}
which is always a stable solution. If $\Delta \alpha$ is non-zero, the equation (\ref{eq:diffPend}) becomes quadratic and the trivial solutions are  
\begin{eqnarray}
	P^*_\pm=\left(f \pm \gamma \right)/\Delta \alpha   \textrm{\ \ \ \ \  for\ \ } \Delta \alpha \neq 0 \label{eq:staticd}\\
	 \textrm{with\ \ }\gamma=\sqrt{\Delta \alpha^2 +2\Delta \alpha \Delta f+ f^2}.
\end{eqnarray}
Only one of these two solutions is stable during propagation along $x$, namely $P^*_-$. Because of $f\geq |\Delta f|$ the solutions $P^*_\pm$ are always real, as expected for a spin polarisation value. This means that for all possible sets of parameters, an equilibrium point can be found.
\subsection{Analytical solution for spin-flip only}

If only spin-flip processes are present, the differential equation (\ref{eq:diffPend}) reduces to 
\begin{eqnarray}
	P'= -fP-\Delta f.
\end{eqnarray}
This equation can be solved by an exponential function with an offset. The general solution takes the form
\begin{eqnarray}
	P=\left(P_0+\frac{\Delta f}{f}\right)\rme^{-fx}-\frac{\Delta f}{f},
	\label{eq:fsolution}
\end{eqnarray}
where $P_0$ is the initial spin polarisation of electrons entering the helical molecules. For large distances, $x \rightarrow \infty$, the spin polarisation $P$ converges to the equilibrium point $P^*$ given by equation (\ref{eq:staticf}). For the special case of only spin independent spin-flips, $\Delta f=0$, one obtains an exponential depolarisation towards $P^*=0$ as a function of film thickness. Such a behaviour was observed by Meier et al. \cite{Meier:1984} for spin transmission through thin films of Ce, Gd, Ni, and Au on a germanium substrate as a dependence of the film thickness. Excitation with circularly polarised light provided an initial spin polarisation $P_0=23.5$\% from the Ge substrate. Meier et al. found mean free paths for spin-flip scattering of 0.32\,nm, 0.38\,nm, 1.25\,nm, and $>$5\,nm for Ce, Gd, Ni, and Au, respectively. The inverse of this mean free path equals the scattering parameter $f$ in our model, in this case with $f=3.1$\,nm$^{-1}$, $2.6$\,nm$^{-1}$, $0.8$\,nm$^{-1}$, and $<0.25$\,nm$^{-1}$ for the respective metals.\\

\subsection{Solution for spin-dependent losses $\Delta \alpha \neq 0$}
It is obvious from transmission experiments in gas phase \cite{Mayer:1995}, electrochemistry \cite{Fontanesi:2017} and photoelectron circular dichroism experiments \cite{Ray:1999,Ray:2006} that spin-dependent electron losses are present in chiral molecules. For the case of a finite spin-dependent loss $\Delta \alpha$ the general solution can be obtained by inserting the two special solutions $P_\pm^*$ according to equation (\ref{eq:staticd}) in the linear reduced Riccati equation (\ref{eq:redRiccati}):
\begin{eqnarray}
	\widetilde{P}'= \mp \gamma \widetilde{P} -\Delta \alpha/2
\end{eqnarray}
where the upper sign is valid for $P_+^*$ and the lower for $P_-^*$, respectively. This equation can be solved for $\gamma\neq 0$ by
\begin{eqnarray}
	\widetilde{P}=C\textrm{e}^{\mp\gamma x}\mp\Delta \alpha/(2\gamma),
\end{eqnarray}
where $C$ is determined by the initial spin polarisation $P_0$ at $x=0$. The spin polarisation $P(x)$ results from equation (\ref{eq:Pubs})
\begin{eqnarray}
	P=\left(f+\gamma\left[\pm1+ \frac{2}{\textrm{e}^{\mp\gamma (x-x_0)}\mp1}\right]\right)/\Delta \alpha
	\label{eq:GenSol}
\end{eqnarray}
where $x_0$ can be obtained from the initial spin polarisation $P_0$. One identifies the term in $[\,]$-brackets as hyperbolic cotangent for the upper sign and hyperbolic tangent functions for the lower sign, so we obtain for the two solutions
\begin{eqnarray}
	P=\left[f-\gamma \coth\left(\frac{\gamma}{2} (x-x_0)\right)\right]/\Delta \alpha
	\label{eq:cothGen}\\
	P={\left[f-\gamma \tanh\left(\frac{\gamma}{2} (x-x_0)\right)\right]}/{\Delta \alpha}.
	\label{eq:tanhGen}
\end{eqnarray}
Obviously, some solutions are not practical for spin polarisations being restricted to values in the interval $[-1, 1] $. With this restriction we have a solution for the spin transport as described by the Riccati equation (\ref{eq:diffPend}). Whether equation (\ref{eq:cothGen}) or (\ref{eq:tanhGen}) solve a specific problem depends on the initial spin polarisation. The hyperbolic cotangent function (\ref{eq:cothGen}) is needed if the initial polarisation $P_0$ is outside of the interval $[P^*_+,P^*_-]$. It always describes a depolarisation. In the case of the spin filter experiments, usually one starts at a low spin polarisation and uses the filter to polarise the electrons. This situation can be described by the hyperbolic tangent function (\ref{eq:tanhGen}). With the boundary conditions $P(0)=P_0$ we get $x_0=2/\gamma \ \tanh^{-1} ([P_0 \Delta \alpha-f]/\gamma)$ for the hyperbolic tangent solution (\ref{eq:tanhGen}) and $x_0=2/\gamma \ \coth^{-1} ([P_0 \Delta \alpha-f]/\gamma)$ in case of the hyperbolic cotangent solution.\\
We now investigate the special case where only loss effects are present with $f=\Delta f=0$ and $\Delta \alpha \neq 0$. Consequently, $\gamma=\Delta \alpha$ and equation (\ref{eq:tanhGen}) reduces to
\begin{eqnarray}
	P=-\tanh \left(\frac{\Delta \alpha}{2} (x-x_0)\right).
\end{eqnarray}
This special case has also been proposed by Farago \cite{Farago:1981} from the Stokes formalism for electrons for a real and diagonal M\"uller matrix. Real and diagonal means that no spin precession or spin-flips are assumed.\\
For completeness we will briefly discuss the case where $\gamma=0$, which was previously excluded. One can show here that the spin-dependent spin-flip parameter and spin-dependent loss parameter must have equal absolute values but opposite sign $\Delta \alpha =-\Delta f$, and additionally that spin-flips can only occur in one direction $f=|\Delta f|$.  The solution in this very special case is $P=\pm 1-2/[\Delta \alpha(x-x_0)],$ with $x_0=2/[\Delta \alpha(P_0\mp1)]$ given by the boundary condition $P_0$, where the upper sign is valid for $\Delta \alpha>0$ and the lower for $\Delta \alpha<0$, respectively. The flip effect and the losses do not fully cancel out and the spin polarisation converges to $P=P^*=\pm 1$ in the direction which is preferred by the spin-flip process.
\subsection{Transmission and circular dichroism}
One can use the results for the spin polarisation in order to calculate transmissions. The solution of equation (\ref{eq:N}) can be obtained by integration
\begin{eqnarray}
	N(x)=N_0\exp\left(-\int^x_0{\frac{\alpha+\Delta \alpha P(x^*)}{2}\mathrm{d}x^*}\right),
\end{eqnarray}
with $N_0$ as the number of electrons at the start at $x=0$. For the solution of interest with $\Delta \alpha\neq 0$, $\gamma\neq 0$ we arrive at
\begin{eqnarray}
	N(x)=N_0\frac{\cosh(\gamma(x-x_0)/2)}{\cosh(\gamma x_0/2)}\ \exp\left(-\frac{\alpha+f}{2}x\right)\label{eq:N(x)} 
\end{eqnarray}
with
\begin{eqnarray}
	 x_0=2/\gamma\ \tanh^{-1} ([P_0 \Delta \alpha-f]/\gamma).\label{eq:x0N(x)}
\end{eqnarray}
This equation describes the number of electrons as a function of the propagation length. The transmission $T$ can be obtained from $T(x)=N(x)/N_0$. Without spin-dependent attenuation  $\Delta \alpha=0$ one gets
\begin{eqnarray}
	N(x)=N_0 \exp\left(-\alpha x/2 \right)
\end{eqnarray}
which is equivalent to the Lambert-Beer law. For completeness we mention the solution in the case of $\gamma =0$:
\begin{eqnarray}
		N(x)=N_0\left(\frac{x}{x_0}-1\right)^{-2}\exp\left(-\frac{\alpha+f}{2}x\right)  \textrm{\ \ with\ \ } x_0=2/(P_0\Delta \alpha-f) \label{eq:bound}.
\end{eqnarray}
In practical experiments it is often hard to determine the absolute electron transmission of a thin film on a substrate. One measures usually relative transmissions according to a change in the initial spin $P_0$ before propagation through the film. The initial spin polarisation can be controlled for example by magnetization of a ferromagnetic substrate or differently circularly polarised radiation causing spin polarised photocurrents from substrates with strong spin-orbit coupling. In such a case one measures the asymmetry $A(x)$ in the photoelectron signal
\begin{eqnarray}
	A(x)=\frac{N^\textrm{\scriptsize{LCP}}(x)-N^\textrm{\scriptsize{RCP}}(x)}{N^\textrm{\scriptsize{LCP}}(x)+N^\textrm{\scriptsize{RCP}}(x)},
	\label{eq:A}
\end{eqnarray}
where $N^\textrm{\scriptsize{LCP}}$ and $N^\textrm{\scriptsize{RCP}}$ are the transmitted electrons excited with left circularly polarised (LCP) light and right circularly polarised (LCP) light, respectively. For a non-chiral measurement geometry, i.e., the light is directed perpendicular onto the sample and no magnetic fields are present, the initial spin polarisation $P_0$ is inverted between the two circular polarisation states of the light ($P_{0}^\textrm{\scriptsize{LCP}}=-P_{0}^\textrm{\scriptsize{RCP}}$). In the most interesting case with $\Delta \alpha\neq 0$, $\gamma\neq 0$ one obtains by inserting equation (\ref{eq:N(x)}) into equation (\ref{eq:A})
\begin{eqnarray}
\fl	A(x)=\frac{\sinh(\tilde{\gamma} x)\sinh(\tilde{\gamma}[x_{0}^\textrm{\scriptsize{RCP}}-x_{0}^\textrm{\scriptsize{LCP}}])}{\cosh(\tilde{\gamma} [x-x_{0}^\textrm{\scriptsize{LCP}}])\cosh(\tilde{\gamma} x_{0}^\textrm{\scriptsize{RCP}})+\cosh(\tilde{\gamma} [x-x_{0}^\textrm{\scriptsize{RCP}}])\cosh(\tilde{\gamma} x_{0}^\textrm{\scriptsize{LCP}} )},
\end{eqnarray}
with $\tilde{\gamma}=\gamma/2$ and the starting condition $x_0^\textrm{\scriptsize{LCP}}$, where $x_0^\textrm{\scriptsize{LCP}}$ is obtained from the respective initial spin polarisation $P_0$ from the substrate by equation (\ref{eq:x0N(x)}). For the special case without spin-flipping ($f=\Delta f=0$) the starting points for LCP and RCP become symmetric $x_{0}^\textrm{\scriptsize{LCP}}=-x_{0}^\textrm{\scriptsize{RCP}}=x_0$ and the asymmetry becomes
\begin{eqnarray}
	A(x)=-P_0^\textrm{\scriptsize{LCP}}\tanh(\Delta \alpha\,x/2)	 \label{eq:NDeltad},
\end{eqnarray}
which is consistent with previous models \cite{Fandreyer:1990,Campbell:1985,Farago:1981}. Without spin-dependent absorption $\Delta \alpha=0$, obviously no asymmetry will be observed, though spin-flip effects may be present.
\section{Comparison with experiments}
In the following we apply this model to data from previously published photoelectron experiments on dsDNA films on Au\cite{Goehler:2011}, oligopeptides on Au \cite{Kettner:2015}, and heptahelicene on Au(111) \cite{Kettner:2018}. In these experiments the spin polarisation of transmitted electrons was measured as a function of the molecular length and thereby the propagation distance $x$.

\subsection{DNA}
\begin{figure}[htbp]
	\centering
		\includegraphics[width=0.7\textwidth]{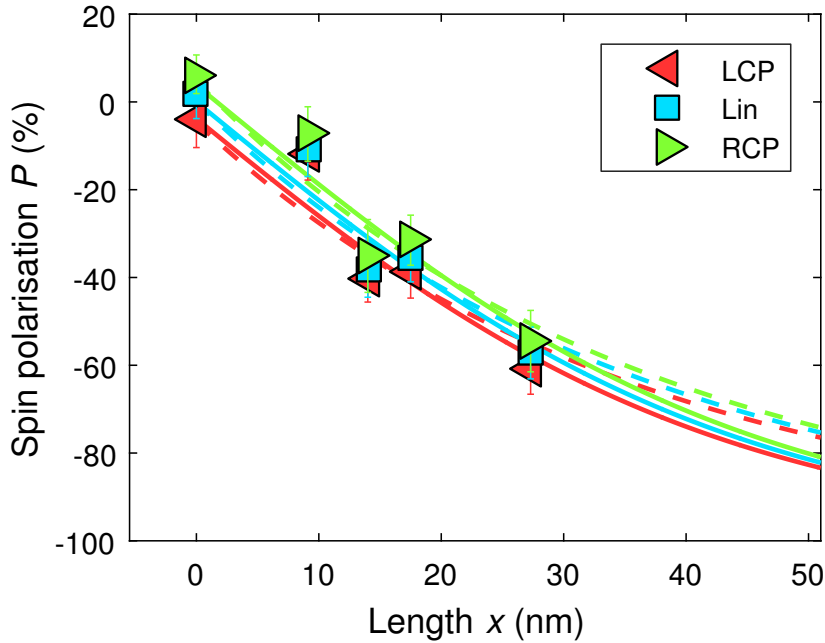}
	\caption{Model fitted to spin polarisations after transmission through DNA molecules with diffrent length \cite{Goehler:2011} (marks). Continuous lines $\Delta f=0$ was set fix, only spin-dependent losses contribute. Dashed lines with boundary condition $\Delta \alpha=0$, only spin-flips contribute.}
	\label{fig:SpinVsXFitDNA}
\end{figure}

G\"ohler et al. \cite{Goehler:2011} measured the spin polarisation of the photocurrent from a polycrystalline gold sample covered with monolayers of double-stranded DNA with various length. In this experiment the initial spin polarisation $P_0$ was varied by changing the polarisation of the exciting UV light. We use a 2D fit routine to model the spin polarisation $P(x,P_0)$ depending on the length of the molecules $x$ and the initial spin polarisation $P_0$. This way one accounts for the full set of 15 data points in a single fit. Figure \ref{fig:SpinVsXFitDNA} depicts the experimental data as marks, while the lines are fitted curves with different boundary conditions. The continuous lines represent a fit under the assumption that only losses are spin-dependent, $\Delta \alpha \neq 0$, while the spin-flip probability for both spin states is equal, i.e., $\Delta f=0$. The fit converges to $\Delta \alpha=(0.046\pm0.02)\,$nm$^{-1}$, $P_0=(-3.8\pm2.3)$\% and no spin-flips $f=0$. The full model including all coefficients $\Delta f$, $f$, $\Delta \alpha$ converges to the same solution within numerical uncertainties and is therefore not shown in Figure \ref{fig:SpinVsXFitDNA}. The dashed lines represent a fit without spin-dependent losses, $\Delta \alpha=0$, but including spin-flip processes. This fit converges to $\Delta f=f=(0.027\pm0.002)\,$nm$^{-1}$. All fits reproduce the data within the accuracy of the measurement \cite{Goehler:2011} of approx $\pm6$\% spin polarisation. Nevertheless, one can see a tendency towards a loss mechanism over a spin-flip mechanism, especially for the data point with the longest DNA molecules. An interesting fact is that all considered fit functions converge to a solution without any depolarising effects. The spin polarisation for an infinitely long molecule converges to a full spin polarisation $P^*=-100$\% in the model. This proposes that CISS can be used to produce perfect spin filters if films with sufficiently long DNA molecules can be prepared. At the length of  $x=51$\,nm, which corresponds to the length of dsDNA in the nucleosome with 147 basepairs, the spin polarisation would reach $P=-82$\%.
\\According to equation (\ref{eq:NDeltad}) one can estimate the asymmetry $A(P_0,\Delta \alpha \cdot x)$  in the photoelectron yield obtained by exciting with circularly polarised light for the cases that only losses determine the spin polarisation. For a length of $x=17.5$\,nm, which corresponds to 50 basepairs, a spin polarisation from the substrate of $P_0=-3.8$\%, and a loss parameter of $\Delta \alpha=0.046$\,nm$^{-1}$ one obtains an asymmetry of $A=(4.2\pm2.6)$\%. In the supplement to ref. \cite{Goehler:2011} an asymmetry of 3.5\% in photoelectron yield was reported for excitation by LCP and RCP light from 50 basepairs (17.5\,nm) long DNA film on polycrystalline gold. The directly measured asymmetry is slightly smaller than the one predicted from fits of the spin measurements but matches well within the experimental accuracy. We expect a trend to a lower asymmetry in the direct measurement, because this circular dichroism experiment was not done with light impinging the sample under normal incidence, but under 60$^\circ$ which may alter the starting polarisation $P_0$ from the substrate.
\subsection{Oligopeptides}
\begin{figure}[htbp]
	\centering
		\includegraphics[width=0.7\textwidth]{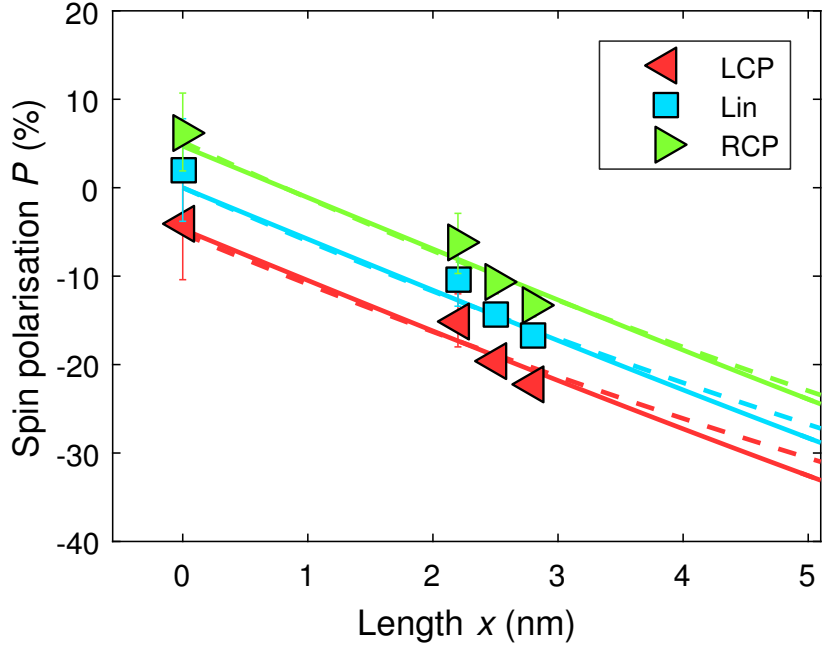}
	\caption{Model fitted to spin polarisations after transmission through oligopeptide molecules with diffrent length \cite{Kettner:2015} (marks). Continuous lines $\Delta f=0$ was set fix, only spin-dependent losses contribute. Dashed lines with boundary condition $\Delta \alpha=0$, only spin-flips contribute.}
	\label{fig:SpinVsXFitOligo}
\end{figure}

Alike the spin measurement of the photo current from DNA coated gold surfaces with different molecular length \cite{Goehler:2011}, measurements from chiral oligopeptides with different length assembled on gold were reported \cite{Kettner:2015}. As one can see in figure \ref{fig:SpinVsXFitOligo}, the spin polarization rises with the molecular length of the oligopeptides. The 2D fitting procedure used in the previous subsection can be applied in this case as well. In the case of a pure spin-flip model, depicted by the dashed lines in figure \ref{fig:SpinVsXFitOligo}, one obtains $\Delta f=0.06$\,nm$^{-1}$ and $f=0.06$\,nm$^{-1}$ with large uncertainties of $\pm0.1$\,nm$^{-1}$, because the spin independent flip parameter $f$ is hard to determine for this dataset including only comparably low absolute spin polarisation values up to about $20$\,\%. In the case with spin-dependent losses and no spin-dependent flips we get $\Delta \alpha=(0.12\pm0.02)$\,nm$^{-1}$ and $f=(0+0.12)$\,nm$^{-1}$ as shown by the solid lines in figure \ref{fig:SpinVsXFitOligo}. Thereby the effect in oligopeptides is about three times stronger per length than in DNA. Though the  low statistics with only three different molecular length and the small absolute spin polarisation values, make it hard to determine more than the approximate strength of the effect from the data.

\subsection{Helicene}
\label{chap:Helicene}
Recently spin-dependent transmissions of electrons from Cu(332), Ag(110) and Au(111) substrates through enantiopure $M$- and $P$-heptahelicene monolayers (($M$)-[7]H , ($P$)-[7]H) were measured in a photoelectron experiment \cite{Kettner:2018}. The change of the spin polarisation by about 8\% is comparably small with respect to DNA, however, [7]H is a much smaller molecule. Furthermore, it seems that the initial spin polarisation from the Au(111) substrate under circularly polarised excitation is only perturbated by an enantiomer specific shift of the spin polarisation after transmission through the [7]H monolayer. This could not be obsevered previously for DNA prepared on Au(111) \cite{Goehler:2011}. However, the results from [7]H on Au(111) allow to quantify the role of spin-flips in CISS with our model. We can check whether the spin polarisation measurements are in agreement with a pure spin-flip model.\\
To approach the problem within the presented model we assume a substrate that inverses the spin polarisation of photoemitted electrons upon altering the circular polarisation state of the light as it is the case for Au(111). We take the difference in spin polarisation of the substrate $\Delta P_0=P_0^\textrm{\scriptsize{LCP}}-P_0^\textrm{\scriptsize{RCP}}$ and compare it with the difference after transmission through a chiral layer $\Delta P_\textrm{\scriptsize{H}}=P_\textrm{\scriptsize{H}}^\textrm{\scriptsize{LCP}}-P_\textrm{\scriptsize{H}}^\textrm{\scriptsize{RCP}}$. If we assume only spin-flip processes we get from equation (\ref{eq:fsolution})
\begin{eqnarray}
	\Delta P_\textrm{\scriptsize{H}}=\Delta P_0\,\, \rme^{-fx}.
\end{eqnarray}
We assumed that the spin polarisation from the substrate is the same for the clean substrate as from the coated substrate before transmission through the film. We can use this to determine
\begin{eqnarray}
	fx=\ln\left(\frac{\Delta P_0}{\Delta P_\textrm{\scriptsize{H}}} \right).
	\label{eq:fhelicene}
\end{eqnarray}
We can determine $\Delta f$ by taking the average of the spin polarisations $\left<P_\textrm{\scriptsize{H}}\right>$ from equation (\ref{eq:fsolution}) from the chiral layer and assume that the spin polarisation from the substrate alone inverts in sign between excitation with LCP and RCP light, $P_0^\textrm{\scriptsize{LCP}}=-P_0^\textrm{\scriptsize{RCP}}$,
\begin{eqnarray}
	\left<P_\textrm{\scriptsize{H}}\right>=\left(P_\textrm{\scriptsize{H}}^\textrm{\scriptsize{LCP}}+P_\textrm{\scriptsize{H}}^\textrm{\scriptsize{RCP}}\right)/2=  \frac{\Delta f}{f}\left(\rme^{-fx}-1\right)\\
	\Delta f=\frac{\left<P_\textrm{\scriptsize{H}}\right>f}{\rme^{-fx}-1}.
	\label{eq:Df}
\end{eqnarray}
We can now straight forward determine $f$ and $\Delta f$. $f_{1,2}$ have to be positive by its definition and thereby $|\Delta f|$ is always smaller than $f$. This inequation,  $|\Delta f|\leq f$, results with equation (\ref{eq:Df}) in
\begin{eqnarray}
	\left|\frac{\left<P_\textrm{\scriptsize{H}}\right>}{\rme^{-fx}-1}\right|\leq 1
\end{eqnarray}
and with equation (\ref{eq:fhelicene}) follows
\begin{eqnarray}
	\left|\frac{\left<P_\textrm{\scriptsize{H}}\right>}{\frac{\Delta P_\textrm{\scriptsize{H}}}{\Delta P_0}-1}\right|\leq 1.
\end{eqnarray}
Because $\Delta P_\textrm{\scriptsize{H}}/\Delta P_0$ is positive and smaller or equal 1 one can conclude
\begin{eqnarray}
	\left|\left<P_\textrm{\scriptsize{H}}\right>\right|\leq 1- \frac{\Delta P_\textrm{\scriptsize{H}}}{\Delta P_0}.
	\label{eq:Ineq}
\end{eqnarray}
If this inequality is not fulfilled, one must include spin-dependent absorption to model the data. The relevant values for the inequality obtained from previously reported measurements \cite{Kettner:2018} are listed in table \ref{tab:Helicene}. One clearly sees that the inequality is violated for ($M$)- and for ($P$)-[7]H. We see that the measurements do not support the assumption that only spin-flip scattering prevails, therefore we must invoke spin-dependent attenuation for the electron transmission through helicene to explain the polarising effect.
\begin{figure}[h]
	\centering
		\includegraphics[width=0.7\textwidth]{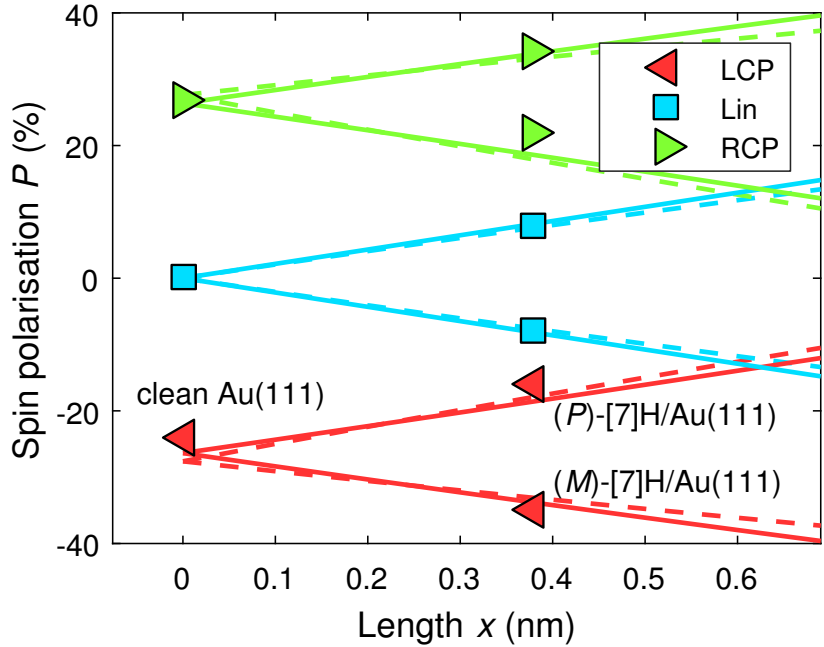}
	\caption{Spin polarisations of the photo currents of Au(111), ($M$)-[7]H/Au(111) and ($P$)-[7]H/Au(111), data (marks) from \cite{Kettner:2018}. The full lines depict a fit with (\ref{eq:tanhGen}) with $\Delta f$ set to zero. The dashed lines depict a fit with only flip processes according to (\ref{eq:fsolution}). Futher details in the text.}
	\label{fig:SpinVsXFitHelicene}
\end{figure}
\\
\begin{table*}[htbp]
	\centering
		\begin{tabular}{|c|c|c|c|c|}
		\hline
			& $P^\textrm{\scriptsize{LCP}}$& $P^\textrm{\scriptsize{RCP}}$& $|\left<P_\textrm{\scriptsize{H}}\right>|$ &  $1- \frac{\Delta P_\textrm{\scriptsize{H}}}{\Delta P_0}$ \\
			\hline
			Au(111)& $-0.242\pm 0.006$ & $0.267\pm 0.005$ &- &-\\
			($M$)-[7]H/Au(111)& $-0.348\pm 0.004$ & $0.218\pm 0.005$ & $0.065\pm0.004$ & $-0.11\pm0.03$ \\
			($P$)-[7]H/Au(111)& $-0.160\pm 0.005$ & $0.341\pm 0.004$ & $0.090\pm0.005$ & $0.015\pm0.04$ \\
			\hline
		\end{tabular}
	\caption{Spin polarisation of photoemitted electrons from [7]H/Au(111) and derived values for inequality (\ref{eq:Ineq}). Data for $P$ taken from \cite{Kettner:2018}. }
	\label{tab:Helicene}
\end{table*}
Analogue to the discussion on the DNA in the previous subsection, we employed a 2D fitting procedure to the spin polarisations from [7]H on Au(111). For symmetry reasons, the values for the spin-dependent parameters $\Delta \alpha$ and $\Delta f$ are inverted for the opposite enantiomer. This enables us to include all data in a single fitting procedure and obtain sufficient statistics. Therefore we virtually inverted the circular polarisation state of the light and the spin for the ($P$) enantiomer. The obtained curves and the spin polarisation data are depicted in figure \ref{fig:SpinVsXFitHelicene}. For the case of the spin-dependent losses and assuming no polarising spin-flips  ($\Delta f=0$) we obtain by fitting equation (\ref{eq:GenSol}) $\Delta \alpha=(0.43\pm0.03)$\,nm$^{-1}$ and $f =(0.00+0.18)$\,nm$^{-1}$. The value $\Delta \alpha$ and thus the strength of the spin filter effect is about ten times larger for [7]H compared to DNA while depolarising spin-flip scattering is low in both cases.

\section{Comparison with theory}
\begin{figure}[htbp]
	\centering
		\includegraphics[width=0.7\textwidth]{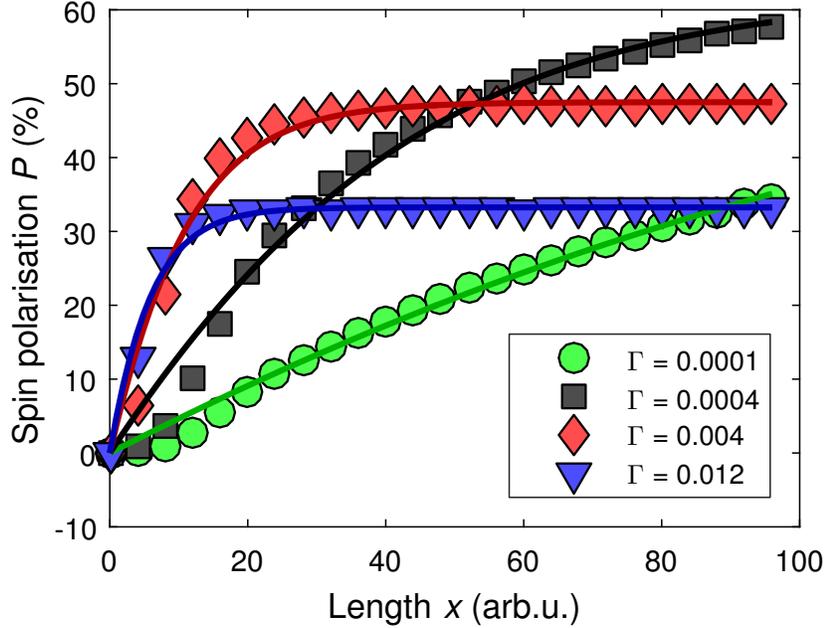}
	\caption{Transport calculations from \cite{Guo:2012b} (marks) fitted by the Riccati like model (\ref{eq:GenSol}) with $\Delta f=0$ (lines)}
	\label{fig:GuoSunFitadscat}
\end{figure}

\begin{figure}[htbp]
	\centering
\includegraphics[width=1\textwidth]{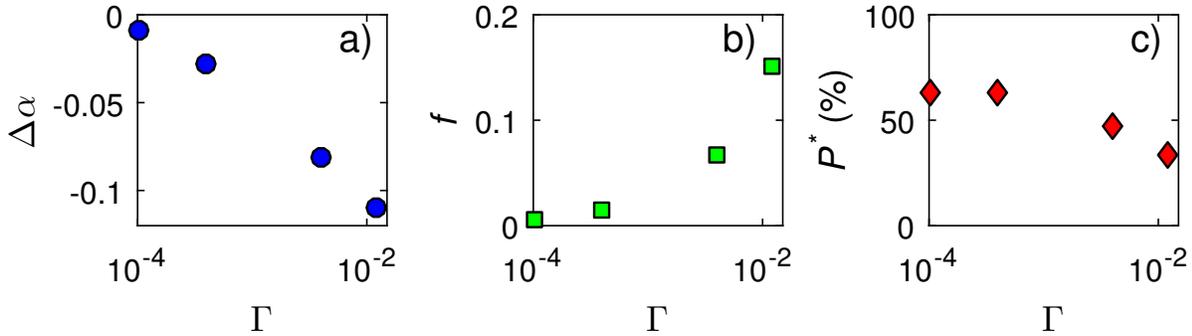} 
\caption{Parameters from fits (lines) in figure \ref{fig:GuoSunFitadscat} as function of the dephasing parameter $\Gamma$ in a semilogarithmic scale. a) spin-dependent losses $\Delta \alpha$, b) spin independent spin scattering $f$, c) maximum spin polarisation $P^*$}
	\label{fig:GuoSunParam}
\end{figure}

The Riccati model can be compared to more sophisticated quantum mechanical models. The marks in the Figure \ref{fig:GuoSunFitadscat} depict numerical results of Sun and Guo \cite{Guo:2012b} for the spinpolarisation $P(x)$ at a fixed electron energy of $E_\textrm{\scriptsize{kin}}=0.486$\,eV. The split step model includes alternating coherent spin-dependent transport and dephasing of the wave functions to model inelastic scattering. The transport is calculated by the Landau-B\"uttiker formalism with a model Hamiltonian for the molecule in form of a double helix dot array. The spin polarisation is increasing with the length of propagation $x$ through the molecule from 0 to up to 30-60\% for the dephasing parameters $\Gamma$ given in the inset. In this model a dephasing is mandatory to obtain a non-zero spin polarisation, on the other hand such a dephasing limits the achievable spin polarisation for long propagation lengths. We use equation (\ref{eq:tanhGen}) to fit $P(x)$ with the boundary condition $\Delta f=0$. The fitted functions are shown as solid lines in figure \ref{fig:GuoSunFitadscat}, and reproduce most features of the numerical data. The parameters obtained from the fits are shown in figure \ref{fig:GuoSunParam} a) and b) as function of the dephasing parameter $\Gamma$. The strength of the filter effect increases from $\Delta \alpha=-0.01$ to $-0.1$ for dephasing parameters between $\Gamma=10^{-4}$ and $0.012$, as depicted in figure \ref{fig:GuoSunParam}a). The spin scattering parameter $f$, figure \ref{fig:GuoSunParam}b), also increases from $f=0.005$ to $0.15$ and as a result the maximum spin polarisation $P^*$ decreases from $P^*=63$\% to $33$\% as depicted in figure \ref{fig:GuoSunParam}c). Consequently, the dephasing $\Gamma$ has a polarising and a depolarising effect on the electron ensemble at the same time. If one allows additional spin-dependent spin-flipping, $\Delta f\neq 0$, the curves hardly change. The slope at the beginning of the spin polarisation as a function of the molecular length can be approximated by the fits in figure \ref{fig:GuoSunFitadscat}, though the inflexion seems slightly stronger in the data from Guo and Sun. The reason for the differences may be that they included inelastic scattering and thus cross talk between different energy channels. Though models with several energy channels may be approached in rate equations too, it is beyond the scope of this article aiming for a most simple description without too many free parameters.

\section{How to find the parameters in an experiment}

\begin{figure}[h]
	\centering
		\includegraphics[width=0.70\textwidth]{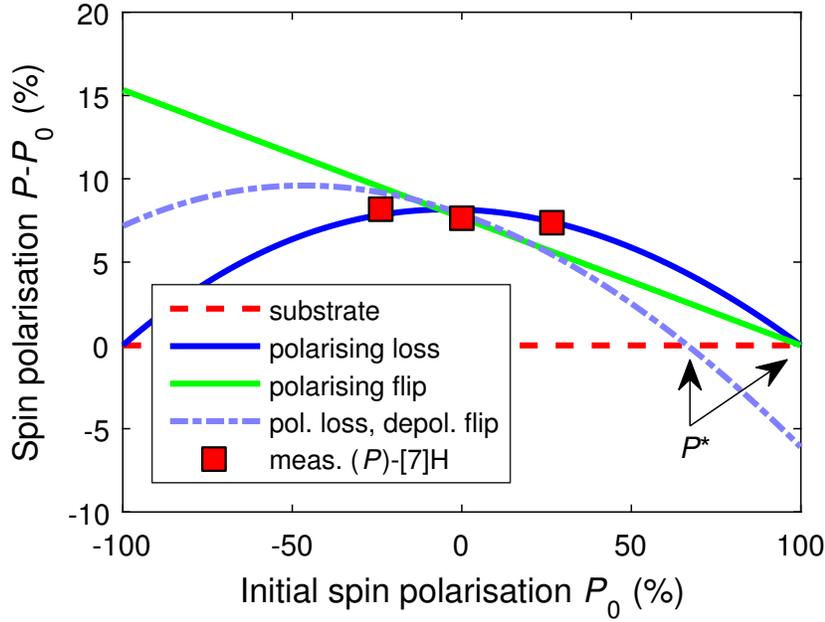}
	\caption{Change in spin polarisation due to transmission through (\textit{P})-heptahelicene layer as function of the spin polarisation from the substrate. Marks: experimental data from \cite{Kettner:2018}.}
	\label{fig:PvsP0}
\end{figure}

\begin{figure}[h]
	\centering
		\includegraphics[width=0.70\textwidth]{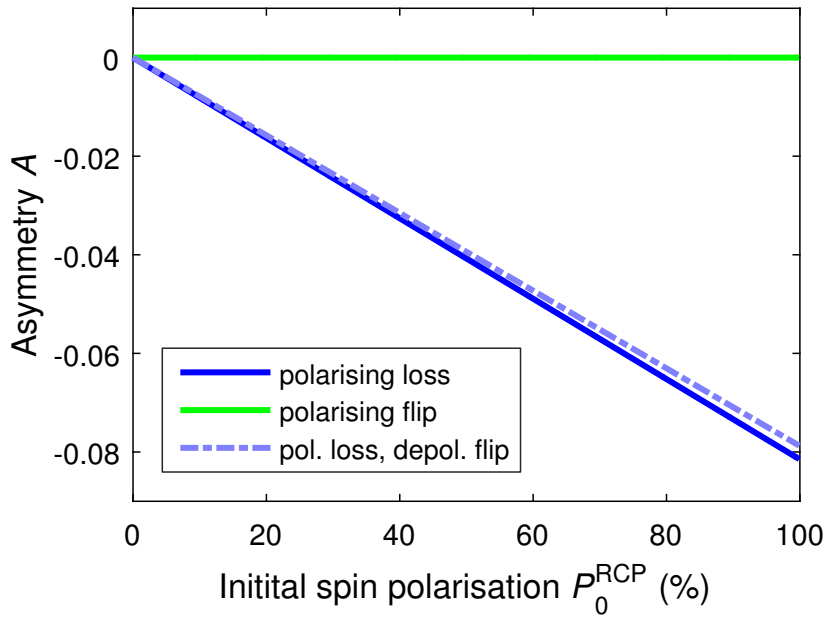}
	\caption{Asymmetry in transmission through (\textit{P})-heptahelicene layer as function of the spin polarisation from the substrate caused by right circularly polarised light, assuming RCP light causes positive spin polarisation.}
	\label{fig:AvsP0}
\end{figure}

As we found out in the previous sections, it is very hard to distinguish between spin-flip and spin-dependent absorption processes by measurements of the electron spin as a function of the molecular length alone. The exponential function (\ref{eq:fsolution}) describes only spin-flips and the more general solution by the hyperbolic tangent (\ref{eq:tanhGen}) is very similar in the range of the experiments, as it can bee seen in figures \ref{fig:SpinVsXFitDNA} and \ref{fig:SpinVsXFitOligo}. Especially around zero spin polarisation both functions are nearly linear.\\
A hint on the kind of processes involved would be to find a saturating spin polarisation with the molecular length in an experiment. In the case of a saturation between $-1<P<1$, spin-flip-processes with $f>\Delta f$ must be involved, otherwise $P=\pm1$ would be reached. A level of saturation could be compared with the stable equilibrium point $P^*$. Such a behavior has not been observed in experiments up to date as it would require very strong spin filters. Another possibility to measure the equilibrium point is to measure the spin polarisation at a fixed molecular length as a function of the initial polarisation. In a photoelectron experiment the initial spin polarisation can be tuned continuously by changing the polarisation of the exciting light from left to right circularly polarised light over intermediate elliptical polarisations. Only at the equilibrium point the final spin polarisation $P(P_0)$ and the initial spin polarisation $P_0$ are equal. This can be used to find the equilibrium point in experimental data.\\
Figure \ref{fig:PvsP0} shows some examples of the difference of the final spin polarisation from the initial spin polarisation as a function of the initial spin polarisation.  The parameters for the calculation have been taken from the evaluation in section \ref{chap:Helicene} on (\textit{P})-heptahelicene which has a length of $x=0.38$\,nm.  For the blue solid curve we assumed only spin dependent losses as in equation (\ref{eq:tanhGen}) with  $\Delta \alpha=-0.43$\,nm$^{-1}$. For the blue dashed curve we invoked an additional depolarising spin-flip parameter with  $f =0.18$\,nm$^{-1}$, which was previously obtained as the maximum for $f$ within one standard deviation. For the green solid curve we assumed a pure spin-flip process according to equation (\ref{eq:fsolution}) with $\Delta f=-f=-0.21$\,nm$^{-1}$. The equilibrium point $P^*$ can be identified as node in the graph at $P_0=100$\% for the solid curves without depolarisation and at $P_0=66$\% for the dashed curve with polarising losses and depolarising spin-flips. From equations (\ref{eq:fsolution}) and (\ref{eq:tanhGen}) one notices a distinct difference in the curves without spin polarising losses and with spin polarising losses. While the function $P(P_0)$ is linear for spin-flips on their own, as soon as spin dependent losses are present, the function becomes nonlinear. The datapoints in figure \ref{fig:PvsP0} are for ($P$)-heptahelicene from the experiment reported in \cite{Kettner:2018}. One clearly sees that the blue solid curve, which describes spin dependent losses, matches the data points best. Thus the most prominent effect in heptahelicene on the electron spin must be spin dependent losses.\\
Another method to check whether the spin polarisation originates from polarising spin-flips or losses, is to record the asymmetry of of the spin polarisation from the chiral film against the initial spin polarisation, as depicted in figure \ref{fig:AvsP0}. This asymmetry is calculated with the same set of parameters used for the previous figure \ref{fig:PvsP0}. The green curve without spin dependent losses is a constant, while with spin dependent losses (blue curves) a linear dependence can be found. For the case without depolarising spin-flips this is expected from the analytical result, namely equation (\ref{eq:NDeltad}). In contrast to the spin polarisation, some additional depolarising spin-flips hardly influence the asymmetry, as can be seen for the blue dashed curve, despite the fact that the spin-flip parameter $f$ reaches already about half the value of the polarising loss parameter $\Delta \alpha$. This means that equation (\ref{eq:NDeltad}) is a good approximation even if some spin-flip scattering takes place and thus $\Delta \alpha$ can be well determined from the slope of the asymmetry.\\
On the other hand this means that experiments measuring only transmission effects are hardly sensitive to depolarizing spin-flips by the parameter $f$ and points out the necessity for direct spin measurements.
The limiting factor in the proposed measurements is the reachable spin polarisation by photoemission with circularly polarised light. With an Au(111) substrate one reaches about 25\% spin polarisation of the photocurrent with photons of an energy of 5.8\,eV \cite{Kettner:2015}. However, very high spin polarised photocurrents have been reached from solid targets. Subashiev et al. \cite{Subashiev:2005} developed a photocathode consisting of a InAlGaAs/GaAsP superlattice structure which produced spin polarised currents with a spin polarisation of up to $P=84$\%. Such a large range for $P_0$ should be more than sufficient to observe or at least accurately interpolate the reachable spin polarisation $P^*$ by the node in the function $P(P_0)-P_0$. According to the results shown in figure \ref{fig:PvsP0}, reducing the initial spin polarization, $(P-P_0)$ allows to clearly distinguish spin dependent flip and loss components in the electron transmission through helical molecules.

\section{Conclusion}
We have set up rate equations for the spin transport in chiral molecules with spin-dependent losses and spin-flip scattering. We derived  from these equations a single differential equation which describes the longitudinal spin polarisation during propagation which is independent of the absolute numbers of spin-up and spin-down electrons. This equation is a Riccati equation, which can be solved fully analytically. The derivations in section \ref{chap:Riccati} are even applicable to non-constant parameters and thus may be used for more complex assemblies. The coefficients in the Riccati equation describe the strength of the filter effect, as well as the probability for spin-flips. From these one does not only gain insight into the origin of the measured spin polarisation, but also obtains upper limits for the spin polarisation for very long molecules. The presented analytical solutions have been applied to discuss experimental data for DNA \cite{Goehler:2011}, oligopeptides \cite{Kettner:2015}, and heptahelicene \cite{Kettner:2018}. From the resulting coefficients one finds that  the smallest molecules, namely [7]H shows the strongest effect with respect to its size.\\
From the spin polarization as a function of the molecular length alone one can hardly distinguish, whether the spin effect is a spin-flipping effect or a spin filter effect, though the fits show a tendency towards a filter effect. We suggest that measurements as a function of the initial spin polarisations before entering the chiral molecules can help to distinguish these cases. We were able to show from such data for [7]H on Au(111) that a spin-flip effect alone can not be held responsible for the measured spin polarisation and also that spin independent spin-flips are comparably unlikely. For a more distinct analysis, comparison with transmission asymmetries under inversion of the initial spin polarisation from the substrate are helpful. The transmission of the chiral molecular film depends on the spin polarisation of the transmitted ensemble as a function of the propagation length. Once this function is known, the transmission can be obtained analytically. The result can be used to obtain the asymmetry in the transmitted electrons under inversion of the initial spin polarisation from the substrate. For DNA on gold, one finds a coinicidence of the measured asymmetry and the direct spin data within a model without spin-flips.\\
As a conclusion, no spin-dependent spin-flips nor spin-independent spin-flips could be detected by our analysis. We found from the analysed experimental data, that spin-independent spin-flips must be much more unlikely compared to the spin-depended absorption, such that high spin polarisations seem feasible to be achieved with long chiral molecules. We propose that the experimental accuracy of these results can be advanced in experiments with large initial spin polarisations from the substrate below the investigated chiral layer.

\ack 
The authors acknowledge financial support from the Volkswagen Stiftung (grant No. 88364).

\section*{Appendix}
\appendix
\setcounter{section}{1}

\bibliographystyle{iopart-num}
\bibliography{paper,books}

\providecommand{\newblock}{}
\begin{thebibliography}{10}
\expandafter\ifx\csname url\endcsname\relax
  \def\url#1{{\tt #1}}\fi
\expandafter\ifx\csname urlprefix\endcsname\relax\def\urlprefix{URL }\fi
\providecommand{\eprint}[2][]{\url{#2}}

\bibitem{Kiran:2016}
Kiran V, Mathew S~P, Cohen S~R, Hern\'andez~Delgado I, Lacour J and Naaman R
  2016 {\em Adv. Mater.\/} {\bf 28} 1957--1962

\bibitem{Kettner:2018}
Kettner M, Maslyuk V~V, N\"{u}renberg D, Seibel J, Gutierrez R, Cuniberti G,
  Ernst K~H and Zacharias H 2018 {\em The Journal of Physical Chemistry
  Letters\/} {\bf 9} 2025--2030

\bibitem{Naaman:2015}
Naaman R and Waldeck D~H 2015 {\em Annu. Rev. Phys. Chem.\/} {\bf 66} 263--281

\bibitem{Mathew:2014}
Mathew S~P, Mondal P~C, Moshe H, Mastai Y and Naaman R 2014 {\em Appl. Phys.
  Lett.\/} {\bf 105} 242408

\bibitem{Koplovitz:2017}
Koplovitz G, Primc D, Ben~Dor O, Yochelis S, Rotem D, Porath D and Paltiel Y
  2017 {\em Adv. Mater.\/}  1606748

\bibitem{Mtangi:2015}
Mtangi W, Kiran V, Fontanesi C and Naaman R 2015 {\em The Journal of Physical
  Chemistry Letters\/} {\bf 6} 4916--4922

\bibitem{Mtangi:2017}
Mtangi W, Tassinari F, Vankayala K, Vargas~Jentzsch A, Adelizzi B, Palmans
  A~R~A, Fontanesi C, Meijer E~W and Naaman R 2017 {\em J. Am. Chem. Soc.\/}
  {\bf 139} 2794--2798

\bibitem{Michaeli:2016}
Michaeli K, Kantor-Uriel N, Naaman R and Waldeck D~H 2016 {\em Chem. Soc.
  Rev.\/} {\bf 45}(23) 6478--6487

\bibitem{Ray:1999}
Ray K, Ananthavel S~P, Waldeck D~H and Naaman R 1999 {\em Science\/} {\bf 283}
  814--816

\bibitem{Ray:2006}
Ray S~G, Daube S~S, Leitus G, Vager Z and Naaman R 2006 {\em Phys. Rev.
  Lett.\/} {\bf 96}(3) 036101

\bibitem{Xie:2011}
Xie Z, Markus T~Z, Cohen S~R, Vager Z, Gutierrez R and Naaman R 2011 {\em Nano
  Lett.\/} {\bf 11} 4652--4655

\bibitem{Goehler:2011}
G\"ohler B, Hamelbeck V, Markus T~Z, Kettner M, Hanne G~F, Vager Z, Naaman R
  and Zacharias H 2011 {\em Science\/} {\bf 331} 894--897

\bibitem{Yeganeh:2009}
Yeganeh S, Ratner M~A, Medina E and Mujica V 2009 {\em The Journal of Chemical
  Physics\/} {\bf 131} 014707

\bibitem{Gutierrez:2012}
Gutierrez R, D\'{i}az E, Naaman R and Cuniberti G 2012 {\em Phys. Rev. B\/}
  {\bf 85}(8) 081404

\bibitem{Guo:2012b}
Guo A~M and Sun Q~f 2012 {\em Phys. Rev. Lett.\/} {\bf 108}(21) 218102

\bibitem{Gutierrez:2013}
Gutierrez R, D\'{i}az E, Gaul C, Brumme T, Dom\'{i}nguez-Adame F and Cuniberti
  G 2013 {\em The Journal of Physical Chemistry C\/} {\bf 117} 22276--22284

\bibitem{Matityahu:2016}
Matityahu S, Utsumi Y, Aharony A, Entin-Wohlman O and Balseiro C~A 2016 {\em
  Phys. Rev. B\/} {\bf 93}(7) 075407

\bibitem{Michaeli:2015}
{Michaeli} K and {Naaman} R 2015 {\em ArXiv\/}
  1512.03435v2[cond--mat.mes--hall]

\bibitem{Michaeli:2017}
Michaeli K, Varade V, Naaman R and Waldeck D~H 2017 {\em J. Phys.: Condens.
  Matter\/} {\bf 29} 103002

\bibitem{Eremko:2013}
Eremko A~A and Loktev V~M 2013 {\em Phys. Rev. B\/} {\bf 88}(16) 165409

\bibitem{Blum:1989}
Blum K and Thompson D 1989 {\em J. Phys. B: At., Mol. Opt. Phys.\/} {\bf 22}
  1823

\bibitem{Medina:2012}
Medina E, L\'opez F, Ratner M~A and Mujica V 2012 {\em EPL (Europhysics
  Letters)\/} {\bf 99} 17006

\bibitem{Rosenberg:2013}
Rosenberg R~A, Symonds J~M, Kalyanaraman V, Markus T, Orlando T~M, Naaman R,
  Medina E~A, L\'opez F~A and Mujica V 2013 {\em The Journal of Physical
  Chemistry C\/} {\bf 117} 22307--22313

\bibitem{Farago:1971}
Farago P~S 1971 {\em Rep. Prog. Phys.\/} {\bf 34} 1055

\bibitem{Fandreyer:1990}
Fandreyer R, Thompson D and Blum K 1990 {\em J. Phys. B: At., Mol. Opt.
  Phys.\/} {\bf 23} 3031

\bibitem{Campbell:1985}
Campbell D~M and Farago P~S 1985 {\em Nature\/} {\bf 318} 52--53

\bibitem{Azzam:1972}
Azzam R~M~A and Bashara N~M 1972 {\em J. Opt. Soc. Am.\/} {\bf 62} 1252--1257

\bibitem{Gardas:2010}
Gardas B 2010 {\em J. Math. Phys.\/} {\bf 51} 062103

\bibitem{BronsteinDeu}
Bronstein I and Semendjajew K 1991 {\em Taschenbuch der Mathematik\/} 25th ed
  (Verlag Harry Deutsch) ISBN 3-87144-492-8 p. 421-422

\bibitem{Meier:1984}
Meier F, Bona G~L and H\"ufner S 1984 {\em Phys. Rev. Lett.\/} {\bf 52}(13)
  1152--1155

\bibitem{Mayer:1995}
Mayer S and Kessler J 1995 {\em Phys. Rev. Lett.\/} {\bf 74}(24) 4803--4806

\bibitem{Fontanesi:2017}
Fontanesi C 2017 {\em Current Opinion in Electrochemistry\/} {\bf 7} 36--41

\bibitem{Farago:1981}
Farago P~S 1981 {\em J. Phys. B: At. Mol. Phys.\/} {\bf 14} L743

\bibitem{Kettner:2015}
Kettner M, G\"ohler B, Zacharias H, Mishra D, Kiran V, Naaman R, Fontanesi C,
  Waldeck D~H, S\c{e}k S, Paw\l{}owski J and Juhaniewicz J 2015 {\em The
  Journal of Physical Chemistry C\/} {\bf 119} 14542--14547

\bibitem{Subashiev:2005}
Subashiev A~V, Gerchikov L~G, Mamaev Y~A, Yashin Y~P, Roberts J~S, Luh D~A,
  Maruyama T and Clendenin J~E 2005 {\em Applied Physics Letters\/} {\bf 86}
  171911

\end{thebibliography}

\end{document}